\documentclass[usenatbib]{mn2e}
\usepackage{graphicx}

\title[A Dearth of Planetary Transits in the direction of NGC~6940]
{A Dearth of Planetary Transits in the direction of NGC~6940}
\author[B. Hood et al.]{Ben Hood$^1$, Andrew Collier Cameron$^1$, Stephen R. Kane$^1$, D.M. Bramich$^{1,}$$^3$,
\newauthor Keith Horne$^1$, Rachel A. Street$^2$, I. A. Bond$^4$, A. J. Penny$^5$,
\newauthor  Y. Tsapras$^6$, A. Quirrenbach$^6$, N. Safizadeh$^7$, D. Mitchell$^7$, J. Cooke$^7$
\\
$^1$School of Physics \& Astronomy, University of St Andrews, North Haugh,
St Andrews, Fife KY16 9SS, Scotland\\
$^2$APS Division, Department of Pure and Applied Physics, Queen's University, Belfast,
University Road, Belfast, BT7 1NN, Northern Ireland\\
$^3$Instituto de Astrofisica de Canarias, C/ Via Lactea s/n, E-38200, La Laguna, Tenerife, Spain\\
$^4$Institute for Astronomy, University of Edinburgh, Royal Observatory, Blackford Hill, Edinburgh, EH9 3HJ, UK\\
$^5$SETI Institute, USA\\
$^6$School of Mathematical Sciences, Queen Mary University of London, Mile End Road, London, E1 4NS, UK\\
$^7$Center for Astrophysics \& Space Sciences (CASS), University of California, San Diego, 9500 Gilman Drive, La Jolla, CA 92093-0424, USA\\
}

\begin{document}

\maketitle

\begin{abstract}

We present results of our survey for planetary transits in the field of NGC 6940. We think nearly all of our observed stars are field stars. We have obtained high precision ($\sim$3-10 millimags at the bright end) photometric observations of $\sim$50,000 stars spanning 18 nights in an attempt to identify low amplitude and short period transit events. We have used a matched filter analysis to identify 14 stars that show multiple events, and four stars that show single transits. Of these 18 candidates, we have identified two that should be further researched. However, none of the candidates are convincing hot Jupiters.

\end{abstract}


\section{Introduction}

\citet{cha00} opened a new chapter in the science of extrasolar planets when they recorded the first transit of a planet around its parent star. The transit produced a 1.5\% dip in the star's light. Until then, the only evidence of planets around main sequence stars had been radial velocity measurements of stellar reflex motions. Though the RV method has been the most successful method of finding planets heretofore, the transit method of searching for planets is complementary, because it provides different information than RV. Measuring a transiting planet can provide the actual mass of the planet by determining the orbital inclination of the system and provide the radius of the planet. Also, in some situations, transiting planets can be probed for atmospheric spectra, as with HD 209458 \citep{bro01}. Finally, the transit method can find planets to kiloparsec distances, much farther than RV.

However, the strength of the transiting method of discovery, that it shows us the orbital inclination, is also its weakness, because that orbital inclination must be close to 90 degrees for us to see the transit. Radial velocity measurements have shown that approximately 1-2\% of Sun-like stars in the solar neighborhood have hot Jupiters, giant planets with orbital distances of 0.035-0.4 AU \citep{lin03}. Assuming that orbital inclinations are random, approximately 10\% of stars with hot Jupiters should have transits visible to us. Therefore, approximately one of a thousand Sun-like stars should show an eclipse, if the stars we observe have the same planetary abundance as the solar neighborhood.

\cite{jan96} suggested that open clusters would be good fields in which to look for planetary transits. Open clusters contain hundreds of stars of a similar distance and metallicity. The field is crowded enough to be able to observe a sufficient number of stars, but not crowded enough to make reduction exceptionally difficult. The high number of stars is essential, since perhaps only one in a thousand stars will exhibit the characteristic dip (a shallow flat-bottomed eclipse) of a planet transiting the parent star. Unfortunately, though this is the reason we observed in the direction of NGC 6940, we don't think we have observed any significant number of cluster members. We describe the reasons for this in more detail in section 2.5 below.

We present results from a deep search for planetary transits in the field of NGC 6940. We describe the observation and data reduction methods used in order to extract light curves for each of these stars. We show that using these methods we can achieve the accuracy necessary to detect planetary transits of a Jupiter-radius object. We describe our transit finding algorithm and show with simulations that we can recover injected transits using that algorithm. Finally, we describe several transit candidates: 14 stars that show multiple low amplitude short duration events and four stars that show single events. We have rejected all but two as poor transit candidates, and recommend them for further study.

\section{Observations and Reduction}
\subsection{Observation}

\begin{figure}
  \includegraphics[angle=0,width=8.2cm]{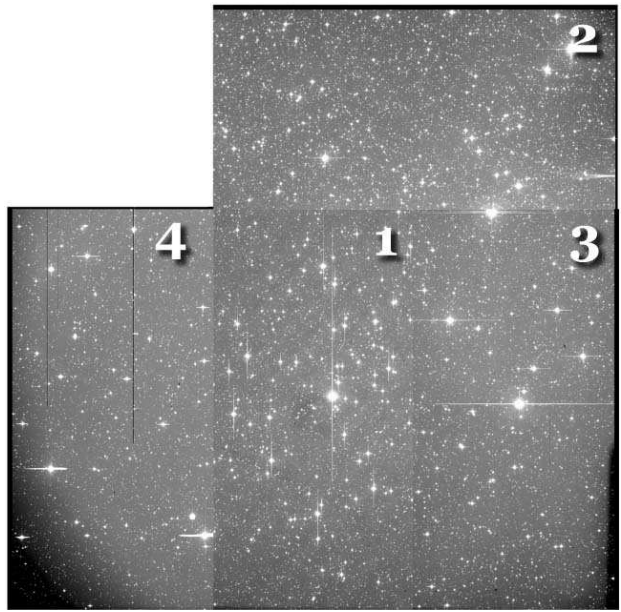}
  \caption{CCD Mosaic of NGC 6940.}
  \label{fig:mosiac}
\end{figure}

Observations were taken over June and July of 1999 using the 2.5 metre Isaac Newton Telescope at La Palma, Canary Islands. Usable observations were taken on 18 nights between 22-30 June and 22-31 July. Images were taken with the Wide Field Camera, a mosaic consisting of four 2048 x 4096 pixel EEV CCDs, mounted at the prime focus of the INT. The mosaic created a 0.29 square degree field of view with 0.33 arcsec per pixel (see Fig. ~\ref{fig:mosiac}).

Three open clusters were observed in rotation during the observing run, NGC 6819 \citep{str03}, NGC 6940, and NGC 7789 \citep{bra04}. This paper reports on the analysis of NGC 6940 (see Table 1) observations. Each image was exposed for 300 seconds, taken in pairs to help remove/identify cosmic rays. This resulted in approximately 2 observations per hour per cluster. We obtained 251, 278, 267 and 249 usable frames of NGC 6940 for each of the four CCDs, respectively. The observing routine was designed to maximize the number of stars observed, in order to maximize the possibility of a transit detection. The 300 second exposure setting was mainly in order to capture enough cluster member stars of NGC 6819 and NGC 7789, which are 1900 and 2400 parsecs distant, respectively. This setting has caused some minor problems with the observation of NGC 6940, discussed below in the section on colours. In retrospect, a shorter exposure time would have been better for NGC 6940, to avoid saturating cluster stars at 770 pc.

\begin{table}
\caption{Parameters of open cluster NGC 6940}
\begin{tabular}{@{}cc}
\hline
RA (J2000.0) & 20 34 26\\
Dec(J2000.0) & +28 17 00\\
$l$ & 69.90\\
$b$ & -7.17\\
Distance(pc) & 770\\
Distance modulus (mag)& 10.10\\
Age($log_{10}$) & 8.858\\
Age(Gyr)& 0.72\\
$[$Fe/H$]$  & +0.01\\
E(B-V) & 0.214\\
\hline
\end{tabular}
\end{table}

\subsection{Data Reduction}

After standard CCD processing, the individual science frames were reduced with differential image analysis, based on code developed by \citet{bon01}. The process is described in more detail by Bramich \citep{bra04} and summarized here. 

We used an automated script and IRAF tools to build a 3-sigma clipped mean masterbias and 3-sigma clipped mean masterflat frame. From each of the science frames we then subtracted the masterbias frame and divided the masterflat. For the reduction procedure, we considered each of the CCDs separately. However, unlike Bramich, we considered all the observations as one run, over June and July 1999, instead of considering them as separate runs.

Following the standard processing, we reduced the photometry on the science frames using differential image analysis (DIA) \citep{ala98, ala00}. Our implementation of DIA code was written for the MOA project \citep{bon01}. All of the processes are automated into scripts which call on C code developed by Bond and Bramich.

Differential Image Analysis (DIA) is excellent for accurately measuring variable stars within a somewhat crowded field. The idea of differential image analysis is that constant stars are removed from the observations, leaving only those stars in which we are interested, because they contain variability induced possibly by a transiting planet. We first used a script to build a reference frame that is a combination of the best seeing frames in the entire run. \citet{ala00} showed that using several good seeing science frames generated better results than just using one, best frame as the reference.

\begin{figure*}
\begin{center}
\begin{tabular}{cc}
  \includegraphics[width=.5\textwidth]{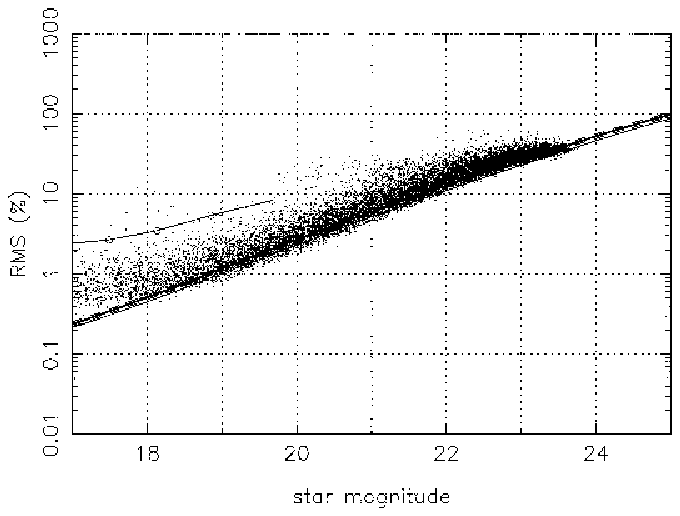}   &
  \hspace{.5cm} 
   \includegraphics[width=.5\textwidth]{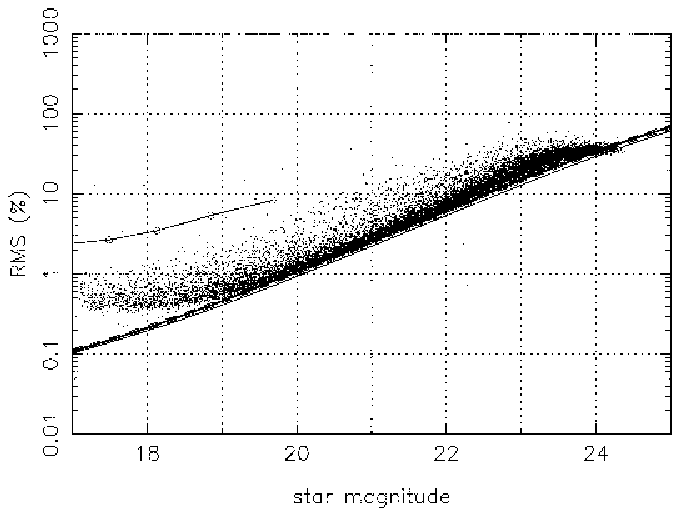} \\
  {(a) CCD 1}&
  {(b) CCD 2}\\
  \includegraphics[width=.5\textwidth]{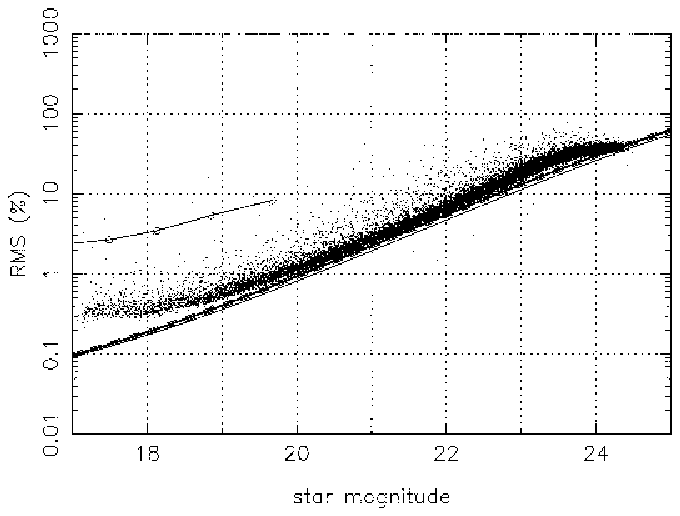}&
   \hspace{.5cm}
  \includegraphics[width=.5\textwidth]{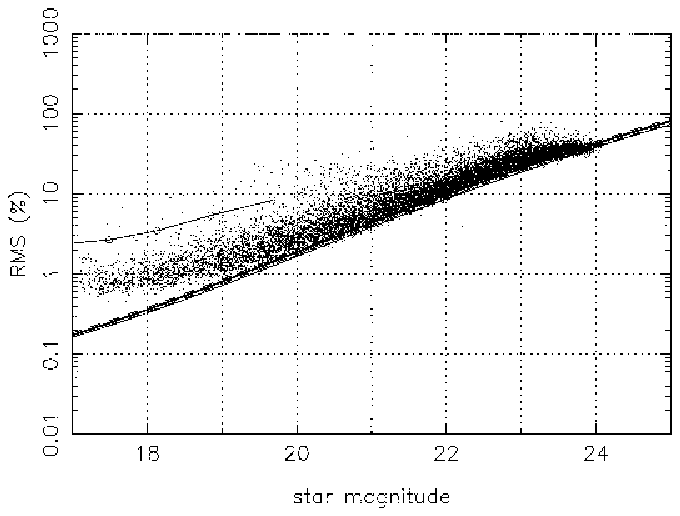}\\
   {(c) CCD 3}&
  {(d) CCD 4}\\
\end{tabular}
\end{center}

\caption{Photometric precision vs R instrumental magnitude.
The lower line represents the theoretical RMS precision based on the CCD noise model. The upper line represents the eclipse depth of a Jupiter sized planet eclipsing a (from left) 0.6, 0.5, 0.4, 0.3 $M_{\sun}$ stars at cluster distance 770 pc.}
\label{fig:pvm}
\end{figure*}

We subtracted this reference frame from each of the science frames to create residual images. In order for the subtraction to be successful, we had to convolve the reference frame to the same seeing as each of the science frames. The science frames $I(x,y)$ are related to the reference frame $R(x,y)$ with the convolution equation:

\begin{equation}
 I(x,y) = K(u,v,x,y) \otimes R(x,y) + B (x,y) \\
 \end{equation}
where $K(u,v,x,y)$ is the convolution kernel and $B(x,y)$ represents the sky background.
Thus, the residual images should have only random noise at the positions of constant stars, while the variable stars will create a dark or light spot on the residual, depending on whether or not the star was dimmer or brighter (relative to the reference frame) in the working image. This method generally performs much better than PSF fitting, particularly with blended stars \citep{ala98}.

Finally, we measure the flux on the residual images using an optimal PSF scaling at the position of each star. Stars have already been identified using PSF fitting on the reference image, using IRAF's DAOPhot package. 

\begin{figure*}
\begin{center}
\begin{tabular}{cc}
  \includegraphics[width=.5\textwidth]{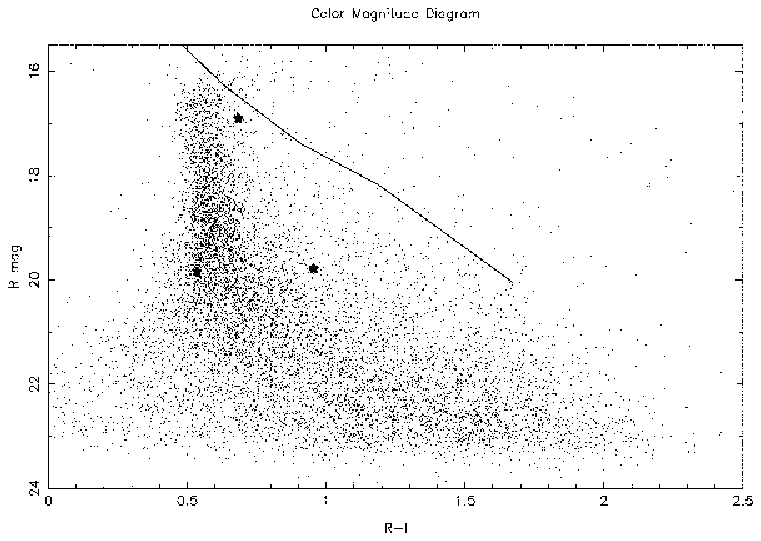}   &
   \includegraphics[width=.5\textwidth]{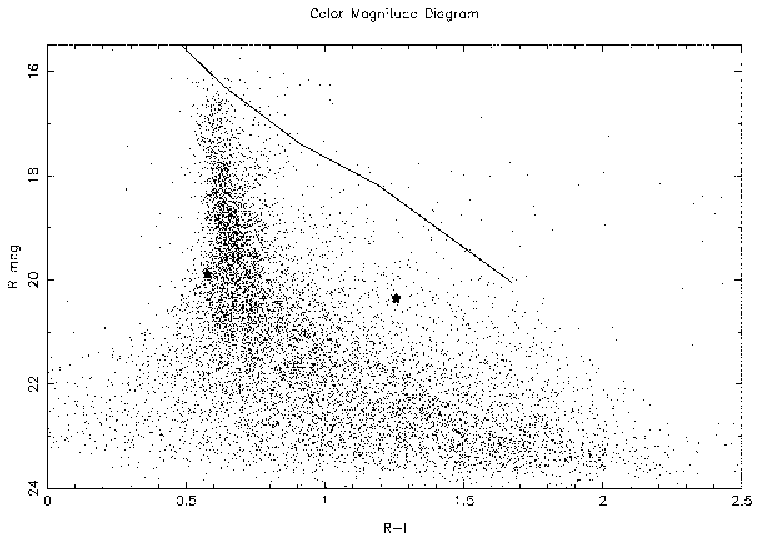} \\
 {(a) CCD 1}&
 {(b) CCD 2}\\
  \includegraphics[width=.5\textwidth]{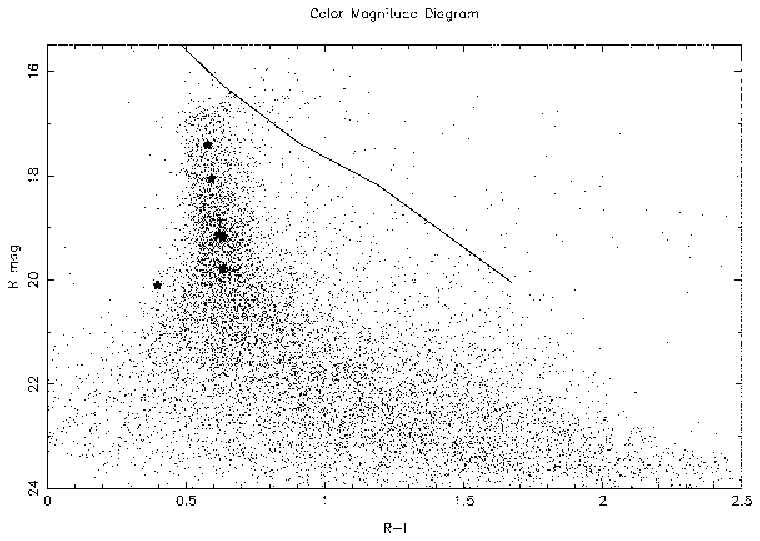}&
  \includegraphics[width=.5\textwidth]{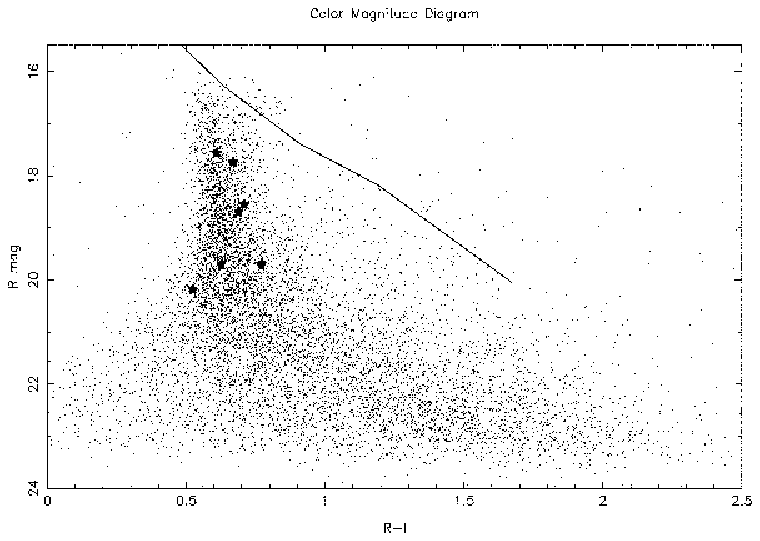}\\
 {(a) CCD 3}&
 {(b) CCD 4}\\
\end{tabular}
\end{center}

  \caption{Colour magnitude diagrams.
Colour magnitude diagrams for each of the four CCDs, with the colours and magnitudes converted to standard values. The highlighted stars are our transit candidates. The line represents a theoretical main sequence for a cluster 770 parsecs away, but only K and M stars would be represented by the line. The main dark ridge in each of the graphs, with $R-I$ colour indices of 0.5-0.75 are spectral type K3-K8.}
\end{figure*}

\subsection {Photometric Precision}
We find that with the above processing, we can achieve an rms scatter of 0.004 -- 0.006 mag at the bright end of our observations; good enough to detect planetary transits (see Fig. \ref{fig:pvm}). However, only a very small number of our stars have precision near this limit. Only $\sim$4400 stars of the $\sim$50,000 have rms scatter better than 1\%.

Our instrumental magnitude saturation limit for each of our CCDs was approximately 17. Beyond that limit, saturated stars, bad columns, and CCD defects were identified as stars. We also see that CCD three (Fig. 2 c) has a much tighter curve than the other three CCDs. This is because we were able to combine 12 best seeing frames in order to make the reference frame for CCD three. The constituent frames of the reference frame need to be roughly sequential, or at least occur on the same night, and only CCD three had such a run of sequential, good seeing frames, without defects. The other CCDs only had four to six sequential frames with good seeing (most had output errors). This created slightly worse reference frames on the other three CCDs and stars with more scatter in the measure of precision.

\begin{figure}
  \includegraphics[width=.5\textwidth]{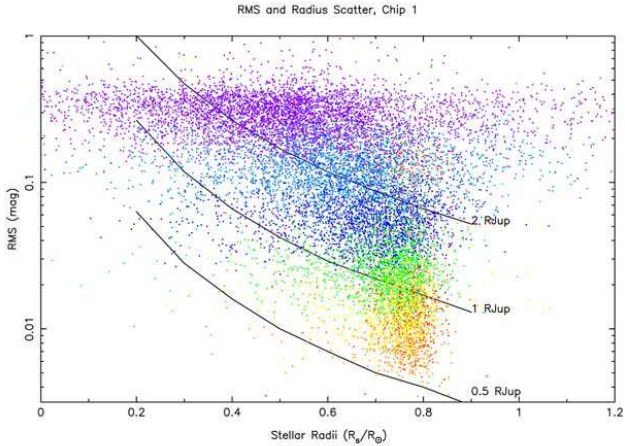}
  \caption{Photometric precision vs stellar radii.
The lines show the transit amplitude that would occur with planets 0.5, 1.0, and 2.0 $R_{Jup}$.}
\end{figure}

\subsection{Colour Data}

We found $\sim$350 photometric standard stars for the field of NGC 6940 from the Canadian Astronomy Data Centre \citep{ste00}. Of these standard stars, we were able to use $\sim$240 stars to calibrate the observations ($\sim$110 of the standard stars were saturated in our data). To change our observations from the instrumental CCD magnitudes into standard $R$ and $I$ magnitudes, we made a linear regression to put CCD one into the standard observations, then corrected each of the other CCDs to conform roughly to CCD one's values.

We began by computing a linear regression between the instrumental $r$ and $i$ values and the standard (Johnson-Cousins) observed $R$ and $I$ values of our 240 stars:
\begin{equation}
R_{\mathrm{JC}} = r_{\mathrm{CCD}}\times0.977+0.192
\end{equation}
and
\begin{equation}
I_{\mathrm{JC}} = i_{\mathrm{CCD}}\times0.985-0.702
\end{equation}

Unfortunately, the photometric standard stars observed were in the center of the cluster, so they only appear on CCD one. Thus, offsets were inferred for the remaining three CCDs by assuming that the mean magnitude in $r$ and the colour $r-i$ of all the stars (to magnitude 20, when we have large errors) would be approximately equal. We found that the following offsets correct the biases of the other CCDs:

\begin{equation}
(r-i)_{\mathrm{CCD2}} = (R-I) + 0.064
\end{equation}

\begin{equation}
r_{\mathrm{CCD2}} = R - 0.105
\end{equation}

\begin{equation}
(r-i)_{\mathrm{CCD3}} = (R-I) - 0.338
\end{equation}

\begin{equation}
r_{\mathrm{CCD3}} = R - 0.177
\end{equation}

\begin{equation}
(r-i)_{\mathrm{CCD4}} = (R-I) - 0.073
\end{equation}

\begin{equation}
r_{\mathrm{CCD4}} = R - 0.148
\end{equation}

\subsection{Colour Magnitude}
Using the calibrations above, we computed the $R-I$ colour index for each star and produced a colour-magnitude diagram (Figure 3). We were unable to find a significant main sequence in the observations of NGC 6940. The 300 s exposures have saturated the members of our cluster, which is only $\sim$770 pc distant, as opposed to the much larger distances of the other clusters ($\sim$1900 and $\sim$2400 pc). Only the K and M cluster members of NGC 6940 were faint enough to be observed, but we believe that not enough of these have been observed to consider our stars part of the cluster. We assume that all our data only refers to field stars. The Besan\c{c}on model for our direction of the galaxy and our observation limits in $R$ estimates that 94.5\% of our observable stars should be K0 spectral type or later \citep{rob03}. This correlates very well with the observed $R-I$ colour indexes for our stars, which suggest that 94.3\% of our stars are of K0 spectral type or later.

\subsection{Stellar Radii}
We used the calibrated $R-I$ index to estimate the stellar radius for each of the stars in our data set. We did this by interpolating between standard values of stellar radius and standard $R-I$ \citep{cox00} to arrive at this polynomial:
\begin{equation}
R_*/R_{\sun}= 1.333-1.548(R-I)+1.131(R-I)^2 -0.3501(R-I)^3
\end{equation}
Judging from the calibrated $R-I$ index, nearly all our main sequence  stars are K and M type. 
We can determine roughly the stars that have sufficient precision by comparing the rms of the star with the depth of a theoretical transit of a Jupiter-sized object. Figure 4 shows the scatter of each of our stars compared to the stellar radii. The lines represent 0.5 $R_{Jup}$, 1 $R_{Jup}$, and 2$R_{Jup}$ transits in front of stars with the appropriate stellar radii. Almost none of our stars have precision good enough to view the transit of a 0.5 $R_{Jup}$ planet, however, about 19\% of our stars have enough precision to measure a one $R_{Jup}$ transit, while nearly 56\% have precision to measure a two $R_{Jup}$ transit.
We have not provided a rigorous treatment of extinction and reddening, which will affect the computed size of the stars, because we are observing field stars, with varying distances.

\section{Transit Detection Algorithm}

The final step in our data reduction is the search for planetary transits from the stellar light curves. We used a matched filter algorithm which compares theoretical transit lightcurves with the observed lightcurves from our $\sim$50,000 reduced stars.

This search uses a truncated cosine approximation with four parameters: period, duration, depth, and the time of transit midpoint. We first used a period sweep from 1.5 d to 7 d with a fixed transit duration of 3 h. The stars with multiple transit-like events are naturally weighted much higher with this method. The fixed-transit duration allows a primary sweep on all stars, which would be too computationally expensive if we varied the duration. A 1.5 $R_{Jup}$ planet with a one day period would create a 1.3--2.0 h transit duration, for stars of spectral type M5-K0. The same planet with a seven day period would create a 2.5--3.8 h duration. We have found that as long as the observed duration does not differ by a factor of two from the fixed duration, our algorithm can identify the transit. 

From this first period sweep, we compute the transit signal-to-noise for each star. The transit S/N is calculated from the fit of the data to a constant light curve as compared to a transit light curve.

Following the first sweep, stars with a significantly better transit fit ($\sim$400 stars, S/N $>$ 8.0) are subjected to another period and duration sweep, which refines the possible transit parameters. Finally, the stars are then analyzed individually (in folded form and unfolded) to consider the possibility of a transit. Stars which have single faint points are rejected, as well as suspicious transits which occur only on nights with known problems.

\section{Detection Simulations}

In order to estimate how many stars might yield planetary transit detections, we used Monte Carlo simulations on two CCDs to estimate how many transit-like events we could recover if every star had a hot-Jupiter sized planet. We ran the simulations on CCDs one and two, and found very similar results. We assume that the other CCDs will show similar results, because all CCDs have similar magnitude distributions.

We began by randomly assigning each star a planetary inclination, planetary period, and planetary transit epoch. The inclinations were uniform in $\cos {i}$, the random period was uniform in $\log p$ from 3 -- 5.2 d, and the epoch of mid-transit was a random date between zero and the period. The planet was assumed to be 1.5 $R_{Jup}$ and the stellar radius was computed using the colour information for each of the stars using equation 10. We then tested each of the systems to determine if the inclination allowed for a transit to be observed, and we compared the transit timing for each of the stars with our actual timings of our observations to see if the simulated transits would occur during our observations. Finally, we injected the transit into the data set using a simple box transit: if an observation was taken during the planet crossing the limb of the star, then the brightness of the star was decreased by half the full transit depth; if it was taken during the full transit, then the magnitude would be offset by the amount computed for a star that size being eclipsed by a 1.5 $R_{Jup}$ planet.

After running this simulation on $\sim$12,500 stars that were recorded on CCD one, we found $\sim$720 stars (5.8\%) that should have transits observable based on inclination and eclipse timing. Similarly, on CCD two we ran 14,000 stars and found $\sim$800 (5.7\%) that would transit. We then inserted these injected transits into our data set and loaded them into {\sc optphot}, our transit search algorithm. We searched over 3 -- 5.2 d periods for 3 h transits. We were able to recover $\sim$370 of the  $\sim$1520 stars with known transits ($\sim$25\%). However, this does not suggest that our algorithm is missing well-defined transits. All stars were given a planet, and over  $\sim$55\% of our stars are magnitude 21 or fainter, with an average precision of 0.05 magnitudes. This precision at faint magnitudes prevents the detection of transits that would only produce shallow dips, especially since it would require many transits during our observing windows, an unlikely event.

We are able to see a distinct differentiation between stars with injected transits and normal observed stars in our transit search. Figure 5 shows that the stars with an injected transit rise significantly above the stars without such a transit. This makes us confident that we would be able to find well defined transits in our brighter stars.

\begin{figure*}
\begin{center}
\begin{tabular}{cc}
  \includegraphics[angle=0,width=.5\textwidth]{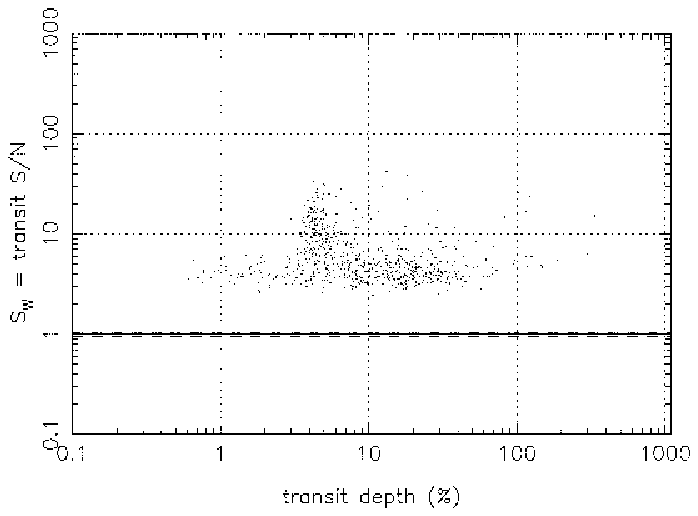} &

  \includegraphics[angle=0,width=.5\textwidth]{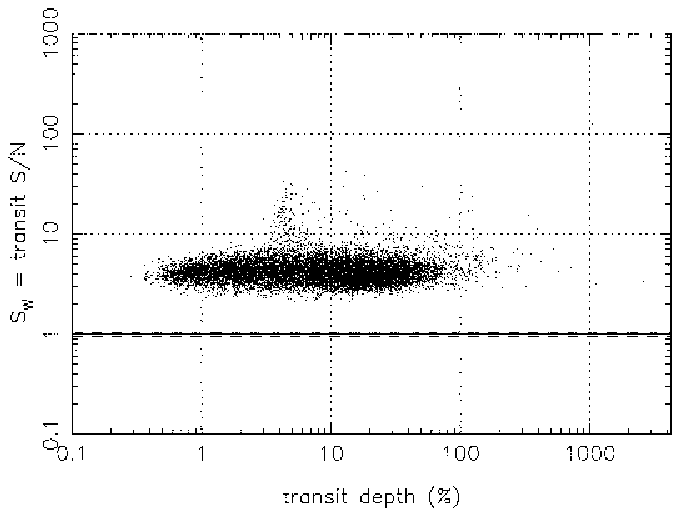} \\
\end{tabular}
\end{center}

\caption{Results of transit injected on CCD one. The left panel shows the results of the $\sim$700 stars which had injected transits. The bulge indicates the easily determined transit signals. The right panel includes all stars on CCD one, including the original stars and stars with an injected transit. Very few stars achieve our cutoff of a S/N of 8.}

\end{figure*}

\section{Results}

Presented in this section are the results from the observations of NGC 6940. Similar results for NGC 7789 or NGC 6819 can be found in \cite{bra04} and \cite{str03}, respectively.

\begin{table*}
\caption{System parameters of stars that show multiple transit-like events. Non-integer values of $N_t$ mean that we observed partial eclipses.}
\begin{tabular}[width=\textwidth]{l|cccccccccccc|}
Star & $R$&$R-I$&$\delta$m& $\delta$t&$R_*$&$R_c$&P\footnote{Our time resolution prevents more accurate period determinations.}&$t_0$ (HJD-&$N_t$&RA&Dec.\\
&(mag)&(mag)&(mag)&(h)&($R_{\sun}$)&($R_{\sun}$)&(d)&2451300)&&(J2000)&(J2000)\\
\hline
\em{CCD 1}\\
6405&16.905(5)&0.692(7)&8.9\%&1.4&0.688(3)&0.205(4)&1.42&51.461&3.5&$20^{\rmn{h}}34^{\rmn{m}}4\fs17$&$+28\degr 09\arcmin 03\farcs79$\\
16016&19.859(6)&0.544(24)&7.2\%&3.0&0.770(15)&0.206(7)&2.17&54.464&4&$20^{\rmn{h}}34^{\rmn{m}}56\fs24$&$+28\degr 17\arcmin 02\farcs48$\\
\em{CCD 2}\\
9939&20.136(9)&1.256(22)&24.4\%&2.1&0.479(8)&0.237(6)&2.20&53.446&2.5&$20^{\rmn{h}}35^{\rmn{m}}13\fs37$&$+28\degr 22\arcmin 11\farcs51$\\
13652&19.891(5)&0.576(21)&17.6\%&4.0&0.750(12)&0.315(6)&2.22&55.499&3&$20^{\rmn{h}}35^{\rmn{m}}27\fs91$&$+28\degr 27\arcmin 38\farcs45$\\
\em{CCD 3}\\
1068&19.127(5)&0.619(13)&8.7\%&3.8&0.725(7)&0.214(5)&7.14&52.465&2&$20^{\rmn{h}}33^{\rmn{m}}35\fs10$&$+28\degr 26\arcmin 53\farcs79$\\
1254&19.793(6)&0.641(23)&18.1\%&2.4&0.714(12)&0.303(6)&4.90&52.597&2&$20^{\rmn{h}}33^{\rmn{m}}36\fs11$&$+28\degr 28\arcmin 34\farcs92$\\
2133&17.413(5)&0.578(9)&3.9\%&3.4&0.749(5)&0.148(10)&3.74&56.769&3&$20^{\rmn{h}}33^{\rmn{m}}41\fs07$&$+28\degr 23\arcmin 29\farcs66$\\
11807&20.104(6)&0.396(36)&11.4\%&2.6&0.876(29)&0.296(10)&5.82&54.513&2&$20^{\rmn{h}}34^{\rmn{m}}34\fs71$&$+28\degr 25\arcmin 09\farcs75$\\
13180&18.062(5)&0.591(10)&7.7\%&2.4&0.741(6)&0.205(5)&4.04&83.596&3&$20^{\rmn{h}}34^{\rmn{m}}41\fs92$&$+28\degr 28\arcmin 17\farcs83$\\
\em{CCD 4}\\
6716&17.564(6)&0.609(10)&12.8\%&4.6&0.731(5)&0.262(6)&3.65&53.628&1.5&$20^{\rmn{h}}34^{\rmn{m}}6\fs28$&$+28\degr 05\arcmin 25\farcs98$\\
7350&17.738(6)&0.669(9)&16.6\%&3.0&0.699(5)&0.284(5)&1.77&55.585&2&$20^{\rmn{h}}34^{\rmn{m}}9\fs78$&$+28\degr 04\arcmin 39\farcs39$\\
8837&18.549(6)&0.708(11)&21.3\%&3.5&0.680(5)&0.314(4)&3.54&53.486&2&$20^{\rmn{h}}34^{\rmn{m}}18\fs74$&$+28\degr 06\arcmin 18\farcs42$\\
12930&20.183(11)&0.523(25)&11.5\%&3.2&0.783(16)&0.265(6)&2.67&51.638&4&$20^{\rmn{h}}34^{\rmn{m}}42\fs05$&$+28\degr 05\arcmin 08\farcs95$\\
15028&18.693(5)&0.689(13)&26.3\%&3.3&0.689(6)&0.353(4)&3.45&57.568&2&$20^{\rmn{h}}34^{\rmn{m}}54\fs87$&$+28\degr 04\arcmin 13\farcs61$\\
\end{tabular}
\end{table*}

\begin{table*}
\caption{System parameters of stars that show single transit-like events}
\begin{tabular}[width=\textwidth]{l|cccccccccc|}
Star & R mag&R-I&$\delta$m& $\delta$t&$R_*$&$R_c$&epoch&RA&Dec.\\
&(mag)&(mag)&(mag)&(h)&($R_{\sun}$)&($R_{\sun}$)&(HJD-2451300)&(J2000.0)&(J2000.0)\\
\hline
\em{CCD 1}\\
1995&19.786(8)&0.956(21)&20.5\%&2.8&0.581(7)&0.263(12)&51.653&$20^{\rmn{h}}33^{\rmn{m}}40\fs32$&$+28\degr 10\arcmin 09\farcs67$\\
\em{CCD 2}\\
\em{CCD 3}\\
11284&19.190(7)&0.631(15)&10.6\%&3.2&0.719(8)&0.234(19)&91.573&$20^{\rmn{h}}34^{\rmn{m}}31\fs69$&$+28\degr 30\arcmin 15\farcs22$\\
\em{CCD 4}\\
1533&19.706(8)&0.626(21)&27.4\%&3.5&0.722(11)&0.378(16)&88.506&$20^{\rmn{h}}33^{\rmn{m}}36\fs75$&$+27\degr 59\arcmin 44\farcs47$\\
2510&19.693(10)&0.771(42)&15.7\%&3.8&0.652(18)&0.258(29)&59.603&$20^{\rmn{h}}33^{\rmn{m}}42\fs35$&$+28\degr 01\arcmin 29\farcs66$\\
\end{tabular}
\end{table*}

\subsection{Multiple Transit-like Events and Variable Stars}

Our transit search algorithm has discovered 14 stars in the field of NGC 6940 that have multiple short duration eclipses. Using the transit depth and stellar radii computed from their colour indices, we have determined a possible radius of each of the stellar companions. Every stellar companion is smaller than 35\% the Sun's radius, and six are smaller than 25\% of the Sun's radius. Folded lightcurves can be found in Fig. 6 while the parameters of each system are found in Table 2. The authors may be contacted for the complete data on each of the candidates to facilitate follow-up work. \\
\begin{figure*}
\begin{center}
\begin{tabular}{ccc}
\includegraphics[angle=0,width=.32\textwidth]{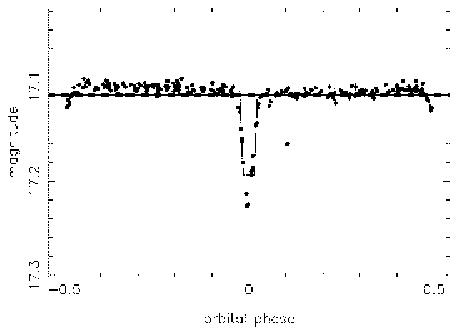}&
\includegraphics[angle=0,width=.32\textwidth]{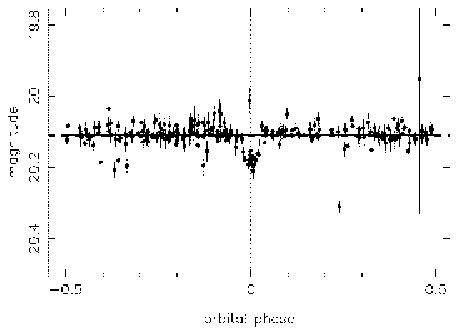}&
\includegraphics[angle=0,width=.32\textwidth]{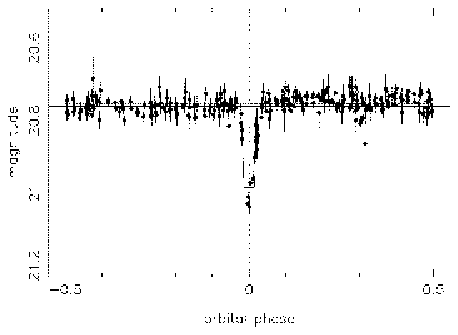} \\
(a) 6405&
(b) 16016&
(c) 9939\\
\includegraphics[angle=0,width=.32\textwidth]{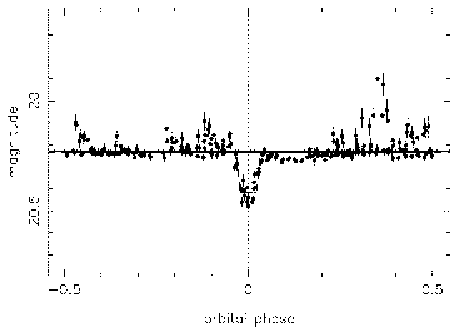}&
\includegraphics[angle=0,width=.32\textwidth]{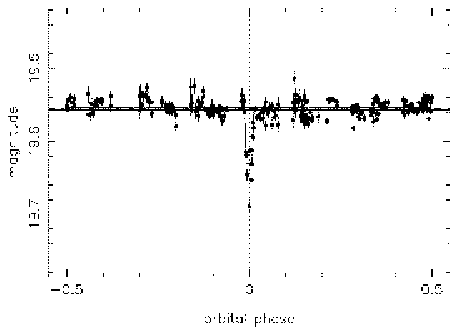}&
\includegraphics[angle=0,width=.32\textwidth]{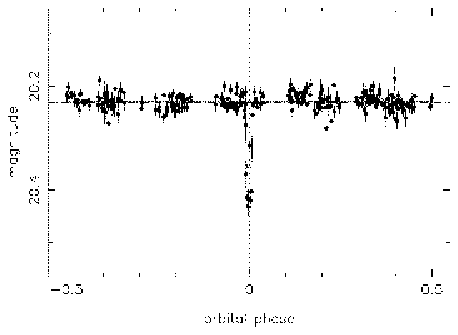}\\
(d) 13652&
(e) 1068&
(f) 1254\\
\includegraphics[angle=0,width=.32\textwidth]{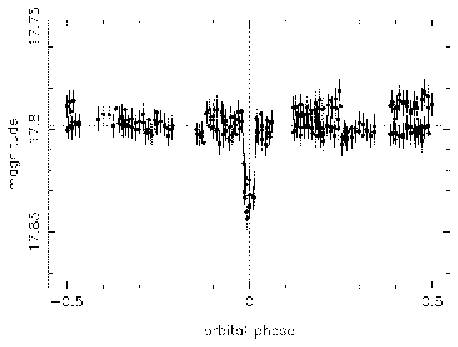}&
\includegraphics[angle=0,width=.32\textwidth]{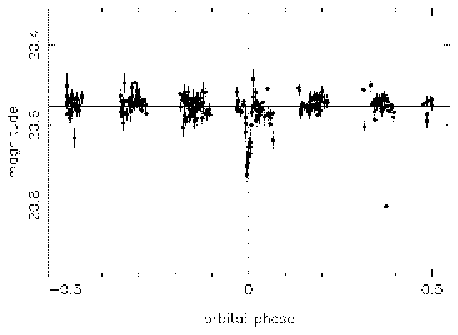}&
\includegraphics[angle=0,width=.32\textwidth]{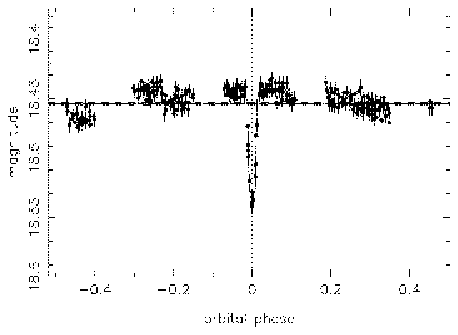}\\
(g) 2133&
(h) 11807&
(i) 13180\\
\includegraphics[angle=0,width=.32\textwidth]{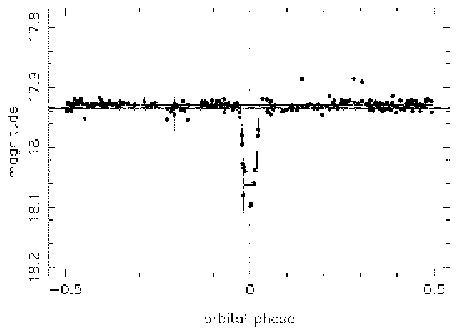}&
\includegraphics[angle=0,width=.32\textwidth]{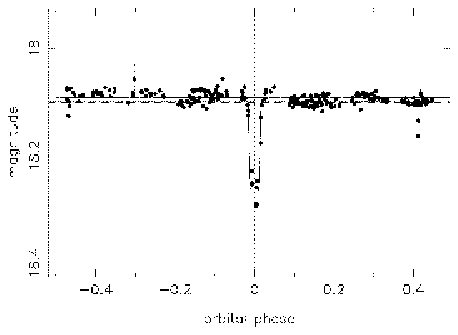}&
\includegraphics[angle=0,width=.32\textwidth]{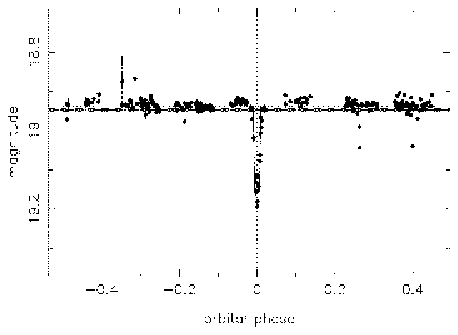}\\
(j) 6716&
(k) 7350&
(l) 8837\\
\includegraphics[angle=0,width=.32\textwidth]{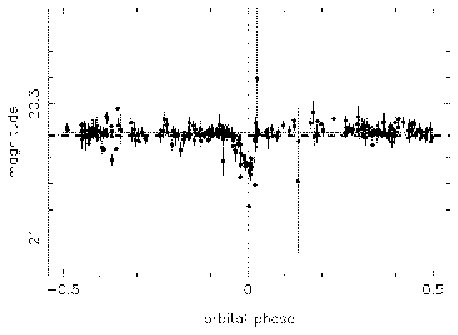}&
\includegraphics[angle=0,width=.32\textwidth]{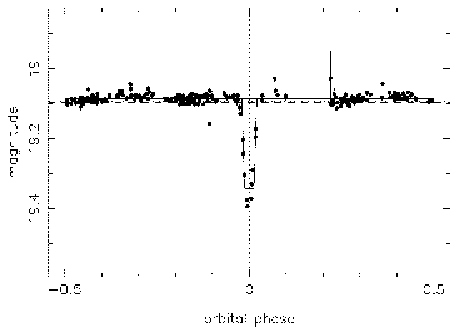}\\
(m) 12930&
(n) 15028\\

\end{tabular}
\end{center}
\caption{Folded lightcurves from each transit candidate. A truncated cosine approximation is used to identify the transit, then all transits are folded together to produce the figures.}
\end{figure*}
\begin{figure*}
\begin{center}
\begin{tabular}{ccc}
\includegraphics[angle=0,width=.32\textwidth]{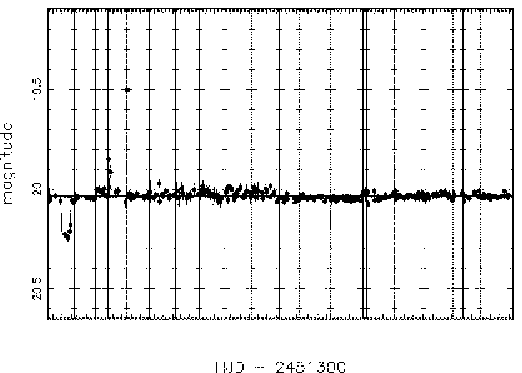}&
\includegraphics[angle=0,width=.32\textwidth]{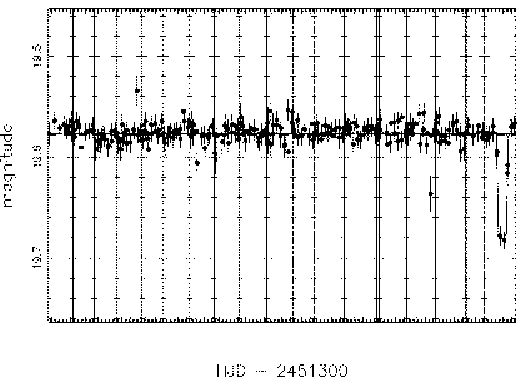}&
\includegraphics[angle=0,width=.32\textwidth]{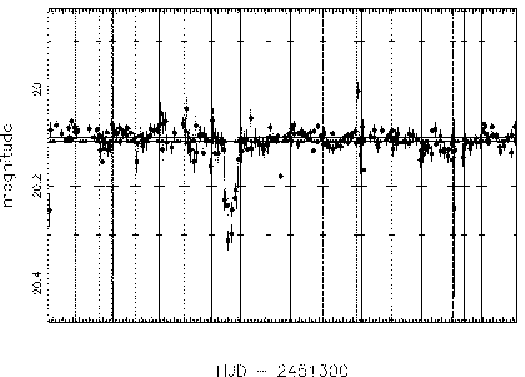}\\
(a) 1995&
(b) 11284&
(c) 2510\\
\includegraphics[angle=0,width=.32\textwidth]{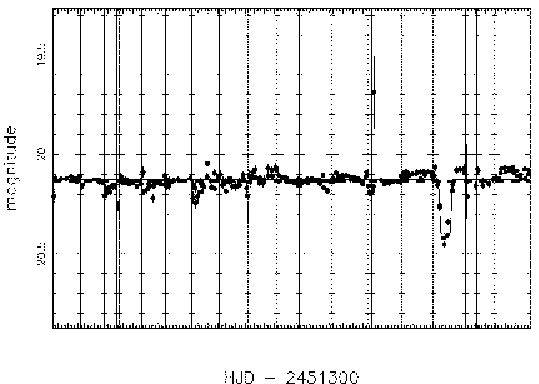}\\
(d) 1533\\
\end{tabular}
\end{center}
\caption{Unfolded lightcurves from single transit events. The vertical lines delineate the nights of observation.}
\end{figure*}
$\mathbf{Star\ 6405:}$ Our eclipse depth of 8.9\% appears to be too conservative, so the assigned companion radius of 2.1 $R_{Jup}$ is probably too small. However, the most damning feature of this eclipse (in terms of it being a planet) is the shallow secondary eclipse that occurs at half the orbital phase. This is definitely a binary system.\\
$\mathbf{Star\ 16016:}$ This is one of our best sampled transits, with 4 transits observed. The noise amplitude of the lightcurve is consistent with other 20th magnitude stars in our sample. The eclipse bottom does not look particularly sharp, though sparse time sampling could have hidden that feature. We have computed a companion radius of 2.1 $R_{Jup}$. If possible, this star should be measured using RV.\\
$\mathbf{Star\ 9939:}$ This has a fairly sharp eclipse, though it is only well sampled on egress, suggesting a grazing binary star. It is also fairly deep, with nearly a 25\% drop in magnitude. However, our colours suggest the parent is an M2 star, with a radius of a little under half a solar radii, giving the companion a radius of $\sim$2.4 $R_{Jup}$.\\
$\mathbf{Star\ 13652:}$ This 18\% eclipse is not as sharp as some of our other obvious binary stars, though the faint magnitude has introduced enough noise to make it difficult to ascertain. The parent star is one of our brighter candidates, a K4. The estimated companion radius is 3.1 $R_{Jup}$, though that is a lower limit, as our eclipse may be deeper than our model suggests. Thus, it is probably a star.\\
$\mathbf{Star\ 1068:}$ If it were indeed a planetary transit, the companion radius would be around 2.2 $R_{Jup}$, orbiting the K5 parent star. However, the sharp eclipse suggests a grazing binary, though sparse time-sampling and few observed eclipses may have contributed to that perception.\\
$\mathbf{Star\ 1254:}$ We cannot really classify if the eclipse is sharp or round bottomed, due to few eclipses and sparse time-sampling, though we think this could be a grazing binary. If the eclipse is caused by a planetary companion to the K5 star, $R_c$ would be around 3 $R_{Jup}$. \\
$\mathbf{Star\ 2133:}$ Out-of-eclipse variation suggests that perhaps this is a binary star. However, it is one of our faintest eclipses, with a 3.9\% dip found with three observed eclipses. This would indicate a planetary companion of 1.5 $R_{Jup}$, which is well within the range for hot Jupiters. We suggest a follow up study of this star. It is also one of the brightest stars in our sample at 17.4 mags, which makes it a good candidate for further research.\\
$\mathbf{Star\ 11807:}$ This faint star seems to have a somewhat sharp eclipse, suggesting a grazing binary. The eclipse depth of 11.4\% may be too conservative, so the computed value of the companion at 3  $R_{Jup}$ is probably too small.\\
$\mathbf{Star\ 13180:}$ This star exhibits some significant out-of-eclipse sinusoidal variation. The sinusoidal period (4.46 days) appears to be slightly but significantly out of sync with the eclipse period (4.04 days). The variation may be star spot activity on the star. A sharp eclipse suggests that this is a binary star grazing its companion.\\
$\mathbf{Star\ 6716:}$Sparse time-sampling prevents us from definitively saying this eclipse has a sharp bottom, but it appears so, suggesting a grazing binary star. The model suggestion of a 12.8\% dip is conservative, so the computed companion size of 2.7 $R_{Jup}$ is a minimum.  \\
$\mathbf{Star\ 7350:}$ This somewhat deep transit could be sharp bottomed, but the time-sampling is too sparse to say for sure. Since the parent is a relatively bright K6 star, we have very little scatter in our data points, and the model fits relatively well.\\
$\mathbf{Star\ 8837:}$ This sharp eclipse has some scatter out of the primary eclipse, and could have a secondary eclipse that we have not yet found. Also, the companion is computed to be larger than 3 $R_{Jup}$, so is probably another star.\\
$\mathbf{Star\ 12930:}$ Though this is one of the faintest stars in our list of candidates, we can find the periodicity because we have luckily observed four transits. However, the scatter does prevent us from saying if the eclipse is sharp or round bottomed.\\
$\mathbf{Star\ 15028:}$ Our models fit this eclipse exceptionally well, but it is fairly deep at 25\%, and suggests a companion radius of 3.6 $R_{Jup}$. Further observations would be necessary to determine the shape of the eclipse.\\

\subsection{Single transit events}
We have also discovered several single low amplitude transit-like events. We are unable to estimate a period for these events, but we can use the colour indices to compute the radii of the stars and the companions. Complete lightcurves are in Fig. 7 and the parameters are found in Table 3. The authors may be contacted for the complete data on each of the candidates.\\
$\mathbf{Star\ 1995:}$ Our models fit this eclipse well within the limited time-sampling, but it is fairly deep. However, our colours indicate a late type star with a radius of 0.581 $R_{\odot}$ which suggests a companion radius of 2.7 $R_{Jup}$.\\
$\mathbf{Star\ 11284:}$ Again, sparse time sampling makes it difficult to characterize the shape of the eclipse. Our computations suggest a companion size of 2.4 $R_{Jup}$. \\
$\mathbf{Star\ 1533:}$ This is a fairly faint star, at nearly 20th magnitude, so there is some amount of scatter in our data points. However, the late type star (K5) can produce this fairly deep eclipse with companion 3.8 $R_{Jup}$, which is a bit large for a planet, and is probably a star. At other points in the data, there could be secondary eclipses that are unresolved with our limited time-sampling, so this could be a faint binary.\\
$\mathbf{Star\ 2510:}$ This single transit eclipse could be sharp-bottomed, and scatter could obscure secondary eclipses. However, we have computed a companion radius of  2.6 $R_{Jup}$.\\

\subsection{System Models: Checking Transit Duration}
Using the stellar parameters in Table 2, we attempted to compare our measured transit duration with a computed transit duration, based on the size of the star (derived from the colour index), the size of the companion (based on the measured transit depth), and the period. Table 4 reports these values of transit durations and the ratio between the two.This is not an absolutely rigorous check (because a missed transit could give us a spurious period), but it does give us some idea as to if the system which we describe is actually a possibility.

We find that four of our stars, 6405, 1068, 1254 and 13180, have observed values within $\sim$20\% of the computed value of the transit duration. We don't feel that this is an endorsement of these candidates as planets (indeed, we know 6405 to be a binary system), but we feel it does probably eliminate the other systems from being planetary systems. Further, each of the four systems have companions computed to be between 2.0--3.0 $R_{Jup}$, too large to be considered planets.

\begin{table}
\caption{Computed versus Observed Transit Duration}
\begin{tabular}[width=.5\textwidth]{ccccc}
star&p (d)&$\Delta t_c$ (h) &$\Delta t_o$ (h)&ratio\\
\hline
6405&1.42&1.1&1.4&0.79\\
16016&2.17&1.5&3.0&0.50\\
9939&2.20&1.3&2.1&0.62\\
13652&2.22&1.6&4.0&0.40\\
1068&7.14&3.2&3.8&0.84\\
1254&4.90&2.7&2.4&1.13\\
2133&3.74&2.0&3.4&0.59\\
11807&5.82&3.3&2.6&1.27\\
13180&4.04&2.2&2.4&0.92\\
6716&3.65&2.2&4.6&0.48\\
7350&1.77&1.3&3.0&0.43\\
8837&3.54&2.2&3.5&0.63\\
12930&2.67&1.8&3.2&0.56\\
15028&3.45&2.2&3.3&0.67\\
\hline
\end{tabular}
\end{table}

\subsection{Modeling the Planet Catch}

Our simulations suggest that if all our stars had a hot Jupiter, $\sim$5.7\% of stars would show an eclipse. That is, the planet's orbit and orbital inclination would allow us to record that transit with our observation regime. Our transit searching algorithm has shown that it can find $\sim$25\% of these transits, if they are randomly distributed over the magnitude ranges we have in our data set. Finally, recent research by \citet{fis04} has quantified the relationship between metallicity and planet frequency, allowing us to quantify how many planets we would expect in our sample, if it mimics the solar neighborhood.

The Besan\c{c}on model supplies us with metallicities for each star in the model, specific to our galactic coordinates. We use these metallicities because we are looking at field stars in the direction of NGC 6940, instead of members of that cluster. We have used the \citet{fis04} data to estimate the probability that each star in our model has a planet, based on its metallicity. We estimate that if the same planet abundance holds in the direction of NGC 6940 as in the solar neighborhood, then $\sim$ 2800 of our stars have Doppler-detectable planets around them, with periods up to three years. Our observation regime and the inclination of the system allow us to see the transit of 5.7\% of those systems, or 160 stars. Further, if the periods of extrasolar planets are assumed to be uniform over log space, then because we are only looking for hot Jupiters, and not planets with periods up to 3 years, we will only see 18\% of the stars with planets, or $\sim$29 transits. Finally, our transit detecting algorithm, when tested on all stars in our data set, was able to find 25\% of transits, or seven transits.

We have produced high precision photometry for $\sim$50000 stars in the direction of NGC 6940. If we use the Besan\c{c}on model coupled with the \citet{fis04} relationships, then we should find about seven hot Jupiters in our data set. However, we have found no convincing hot Jupiters. Nearly all of our `candidates' are almost certainly grazing binary stars, though a few simply have too little data to define them. One of our stars (16016) has an eclipse that might be round bottomed, and the computed radius of the companion is 2.1 $R_{Jup}$, which may be an M dwarf. Another (2133), which exhibits out of eclipse variation (suggesting a binary star) has a computed companion radius of 1.5 $R_{Jup}$, which is well within the range of hot Jupiter radii. We recommend these stars for further study.

We can use Poisson statistics to estimate the significance of our null result, using:\\
\begin{equation}
f(x)=\frac{e^{-a} a^x}{x!}
\end{equation}
where $a$ is our expected number of planets (seven) and $x$ is our actual planet catch (zero). We use this to calculate that there is only a $9.12\times10^{-4}$ chance of finding zero planets when we expect seven. This gives us a $3.3\sigma$ null significance.

The lack of detections is surprising, even given the expected metallicity distribution in the stellar field population we surveyed. The main systematic difference between the population studied here and the solar neighborhood samples studied by \citet{fis04} is that our stars are predominantly late K or M dwarfs of 0.7 $R_{\sun}$ or smaller. Radial velocity surveys have only discovered two M dwarfs harbouring planets, but that could be an observational bias against M dwarfs, which are often too faint for RV studies. Our results point to a lower incidence of hot Jupiters among late K and M dwarfs than among F or G dwarfs, regardless of the metallicity. \citet{end03} have embarked on a study specifically aimed at finding if the formation history of M dwarfs prevents planetary companions, though they have not finished their surveys. Our results thus suggest that hot Jupiters are less common around M dwarfs, and the lack of planets is not an observational bias.

\section{Conclusions}

We have obtained high precision light curves for $\sim$50,000 stars in the direction of the open cluster NGC 6940 using differential image analysis. We have used Monte Carlo simulations to estimate how many transiting planets we should expect to find, assuming planetary frequency of the solar neighborhood. We determined the sizes of the stars using colour information from our observations and calibrated stars from \citet{ste00}. Using a matched filter algorithm, we have identified several stars that exhibit behavior similar to that which is produced by an extrasolar planet. However, most of our candidates exhibit secondary or sharp-bottomed eclipses, suggesting that the stars in question are binary stars and not stellar systems with hot Jupiters.

We have been unable to find the number of stars with transiting planets we estimated we would find. This could be because we are looking at mostly late type K and M stars, instead of earlier type F and G stars. We have identified several candidates with multiple transit-like events, and some with single events, though none are unambiguously caused by planetary companions.

\section*{Acknowledgements}

Ben Hood would like to thank the Marshall Commission for financial support. The authors would also like to thank Aleksander Schwarzenberg-Czerny for useful discussions regarding the transit detection algorithm. This research was (partially) based on data from the ING Archive. We thank the Canadian Astronomy Data Centre, which is operated by the Dominion Astrophysical Observatory for the National Research Council of Canada's Herzberg Institute of Astrophysics. This paper was based on observations made with the Isaac Newton Telescope operated on the island of La Palma by the Isaac Newton Group in the Spanish Observatorio del Roque de los Muchachos of the Instituto de Astrofisica de Canarias.

\end{document}